\documentclass{elsart}
 \usepackage{graphicx}%USES EPS FIGURES!!!!!!!!!!!!
 \usepackage{amssymb}
 \usepackage{lineno}

\def\beq{\begin{equation}}
\def\eeq{\end{equation}}
\def\bea{\begin{eqnarray}}
\def\eea{\end{eqnarray}}
 \def\r{{\bf r}}
 
\def\p{{\bf p}}
 
 \def\n{{\bf n}}
 \def\I{{\bf I}}
 \def\half{\frac{1}{2}}

 \def\MeV{{\rm MeV}}
 \def\half{{\frac{1}{2}}}
\begin{document}
 \begin{frontmatter}
\title{Kinetic equation for finite systems of fermions\\
 with pairing }%con correzioni di Ciccio e di David
\author[label1]{ V. I. Abrosimov},
\author[label2]{D. M. Brink},
\author[label3]{A. Dellafiore\corauthref{cor1}},
\corauth[cor1]{Corresponding author}
\ead{della@fi.infn.it}
\author[label3,label4]{F. Matera}
\address[label1]{Institute for Nuclear Research, 03028 Kiev, Ukraine}
\address[label2]{Oxford University, Oxford, U.K.}
\address[label3]{Istituto Nazionale di Fisica Nucleare, Sezione di Firenze}
\address[label4]{Dipartimento di Fisica, Universit\`a degli Studi di Firenze, via Sansone 1,  I 50019 Sesto F.no  (Firenze), Italy}

\begin{abstract}
The solutions of the Wigner-transformed time-dependent Hartree--Fock--Bogoliubov equations are studied in the constant-$\Delta$ approximation.  This approximation is known to violate  particle-number conservation. As a consequence, the density fluctuation and the longitudinal response function given by this approximation contain spurious contributions.  A simple prescription for restoring both local and global particle-number conservation is proposed.  Explicit expressions for the eigenfrequencies of the correlated systems and for the density response function are derived and  it is shown that the semiclassical  analogous of the quantum single--particle spectrum has an excitation gap of $2\Delta$, in agreement with the  quantum result.  The collective response is studied for a simplified form of the residual interaction.
\end{abstract}
\begin{keyword}
Pairing \sep Vlasov equation 
\PACS 21.10.Pc \sep 03.65.Sq
\end{keyword}
\end{frontmatter}
\section{Introduction}
The problem of extending the Vlasov equation to systems in which pairing correlations play an important role has been tackled some time ago by Di Toro and Kolomietz \cite{dk} in a nuclear physics context and, more recently, by Urban and Schuck \cite{us} for trapped fermion droplets. These last authors derived the  TDHFB equations for the Wigner transform of the normal  density matrix  $\rho$ and of the pair correlation function $\kappa$ (plus their  time-reversal conjugates) and used them to study the dynamics of a spin-saturated trapped Fermi gas. In the time-dependent theory  one obtains a system of four coupled differential equations  for $\rho$, $\kappa$, and their  conjugates \cite{us} and, if one wants an  analytical solution, some approximation must be introduced.  Here we try to find a solution of the equations of motion derived by  Urban and Schuck  in the approximation in which the pairing field $\Delta(\r,\p,t)$ is treated as a constant. It is well known that such an approximation violates both particle-number-conservation and gauge invariance (see e.g. sect. 8-5 of \cite{sch} and \cite{cks}), nonetheless we study it because of its simplicity, with the aim of correcting the final results for its shortcomings. 
Moreover, the constant-$\Delta$ approximation is not satisfactory for describing long
wavelength pairing modes in a large system. Such modes have frequencies which
are much less than the pairing frequency $\Delta/\hbar$ and for their study it is essential to
use a self consistent theory where the gap $\Delta$ is related to the pair
density $\kappa$ through the pairing interaction. The phases of $\Delta$ and
$\kappa$ are particularly important because they describe the collective
superfluid currents. On the other hand nuclei are small systems. Shell gaps
are large compared with $\Delta$, or equivalently giant resonance frequencies
are large compared with the pairing frequency. The constant-$\Delta$
approximation is much more reasonable in such systems.

In Sect. 2, the basic equations are recalled and reformulated in terms of the even and odd components of the normal density $\rho$. In Sect. 3, the static limit is studied by following the approach of \cite{bs} and the constant-$\Delta$ approximation is introduced. In Sect. 4, the simplified dynamic equations resulting from the constant-$\Delta$ approxmation are derived and their solutions are determined in linear approximation. In Sect. 5, these solutions are studied in a one-dimensional model and the problem of particle-number conservation is examined in detail. By studying the energy-weighted sum rule (in the Appendix), we find that the constant-$\Delta$ approximation introduces some spurious strength into the density response of the system. A simple prescription, based on the continuity equation, is proposed in order to eliminate the spurious strength. The resulting strength function gives the same energy-weighted sum rule as for the uncorrelated systems.  In Sect. 6, the general solution found in Sect. 4 is re-written for spherical systems, where the angular integrations can be performed explicitly, leading to expressions containing only radial integrations. In Sect. 7, the collective response function  of spherical nuclei is derived for
 a simple multipole-multipole residual interaction.  In Sect. 8, the quadrupole and octupole channels, that are the ones most affected by the pairing correlations are shown explicitly. Finally, in Sect. 9 conclusions are drawn.

\section{Basic equations}
We assume that our system is saturated both in spin and  isospin space and do not distinguish between neutrons and protons, so  we can use directly the equations of motion of Urban and Schuck.

We start from the equations of motion derived in Ref. \cite{us} for the Wigner-transformed density matrices $\rho=\rho_(\r,\p,t)$ and $\kappa=\kappa(\r,\p,t)$, with the warning that the sign of $\kappa$ that we are using agrees with that of Ref. \cite{dk}, hence it is opposite to that of \cite{us}. 
Moreover we find convenient to use the odd and even combinations of the normal density introduced in \cite{us}:
\bea
\label{roev}
&&\rho_{ev}=\half[\rho(\r,\p,t)+\rho(\r,-\p,t)]\,,\\
\label{rood}
&&\rho_{od}=\half[\rho(\r,\p,t)-\rho(\r,-\p,t)]\,.
\eea

 Thus, the equations of motion given by Eqs.(15a...d) of Ref. \cite{us} read
\bea
\label{even}
i\hbar\partial_t\rho_{ev}&=&i\hbar\{h,\rho_{od}\}-2i{\rm Im}[\Delta^*(\r,\p,t)\kappa]\,\\
\label{odd}
i\hbar\partial_t\rho_{od}&=&i\hbar\{h,\rho_{ev}\}+i\hbar{\rm Re}\{\Delta^*(\r,\p,t),\kappa\}\,\\
\label{kappa}
i\hbar\partial_t\kappa&=&2(h-\mu)\kappa-\Delta(\r,\p,t)(2\rho_{ev}-1)+i\hbar\{\Delta(\r,\p,t),\rho_{od}\}\,.
\eea
Here $h$ is the Wigner-transformed Hartree--Fock hamiltonian $h(\r,\p,t)$, while $\Delta(\r,\p,t)$ is the Wigner-transformed pairing field.
Since the time-dependent part of $\kappa$ is complex, $\kappa=\kappa_r+i\kappa_i$,  the last equation gives two separate equations for the real and imaginary parts of $\kappa$.

Moreover, from the supplementary normalization condition (\cite{rs}, p. 252)
\beq
{\cal R}^2= {\cal R}\,
\eeq
satisfied by the generalized density matrix ${\cal R}$, the two following independent equations are obtained:
\bea
\label{scpau1}
\rho_{od}\kappa+i\frac{\hbar}{2}\{\rho_{ev},\kappa\}&=&0\,,\\
\label{scpau2}
\rho_{ev}(\rho_{ev}-1)+\rho_{od}^2+\kappa\kappa^*&=&0\,.
\eea
We shall use  the equations of motion (\ref{even}--\ref{kappa}), together with these equations, as our starting point, but
first  we  notice that, in the limit of no pairing, both $\Delta$ and $\kappa$ vanish, the third equation of motion reduces to a trivial identity, while the first two give the Vlasov equation for normal systems, expressed in terms of the  even and odd components of $\rho$:
\bea
\label{vla1}
\partial_t\rho_{ev}&=&\{h,\rho_{od}\}\,,\\
\label{vla2}
\partial_t\rho_{od}&=&\{h,\rho_{ev}\}\,.
\eea
A solution of the linearized Vlasov equation for normal systems (i. e. without pairing) has been obtained in Ref. \cite{bdd} and our aim here is to study the changes introduced by the pairing interaction in the solution of \cite{bdd}.

Moreover, before studying the time-dependent problem, it is useful to look at the static limit.

\section{Static limit}

In this section we follow the approach of Ref. \cite{bs}.
At equilibrium we have
\bea
\rho_{ev}&=&\rho_0(\r,\p)\,,\\
\rho_{od}&=&0\,,\\
h&=&h_0(\r,\p)\,,\\
\kappa&=&\kappa_0(\r,\p)\,\\
\Delta_0&=&\Delta_0(\r,\p)
\eea
and  equations (\ref{even}--\ref{kappa})  give
\bea
\label{even2}
0&=&-2i{\rm Im}(\Delta_0^*\kappa_0)\,,\\
\label{odd2}
0&=&i\hbar\{h_0,\rho_0\}+i\hbar{\rm Re}\{\Delta_0^*,\kappa_0\}\,,\\
\label{kappa2}
0&=&2(h_0-\mu)\kappa_0-\Delta_0(2\rho_0-1)\,,
\eea
while Eqs. (\ref{scpau1}, \ref{scpau2}) give
\bea
\label{stpau1}
i\frac{\hbar}{2}\{\rho_0,\kappa_0\}&=&0\,,\\
\label{stpau2}
\rho_{0}(\rho_{0}-1)+|\kappa_0|^2&=&0\,.
\eea

Equation (\ref{even2}) is  satisfied if we assume that $\Delta_0$ and $\kappa_0$ are real quantities, while Eqs. (\ref{kappa2}) and (\ref{stpau2}), taken as a system,  have the solution:\cite{bs}

\bea
\label{roz}
&&\rho_{0}(\r,\p)=\frac{1}{2}{\Big(}1-\frac{h_0(\r,\p)-\mu}{E(\r,\p)}{\Big)}\,\\
&&\kappa_0(\r,\p)=-\frac{\Delta_0(\r,\p)}{2E(\r,\p)}\,,
\label{kappaz}
\eea
with the quasiparticle energy
\beq
\label{qpe}
E(\r,\p)=\sqrt{\Delta_0^2(\r,\p)+(h_0(\r,\p)-\mu)^2}\,.
\eeq

It can be easily checked that Eqs. (\ref{roz}, \ref{kappaz}) satisfy also Eqs. (\ref{odd2}) and (\ref{stpau1}), that is
\beq
\{h_0,\rho_0\}+\{\Delta_0,\kappa_0\}=0\,
\eeq
and
\beq
\{\rho_0,\kappa_0\}=0\,.
\eeq

The (semi)classical equilibrium phase-space distribution is closely related to $\rho_0(\r,\p)$:
\beq
f_0(\r,\p)=\frac{4}{(2\pi\hbar)^3}\rho_0(\r,\p)
\eeq
and the statistical factor $4$ takes into account the fact that there are two kinds of fermions.

The parametrer $\mu$ is determined by the condition
\beq
\label{norm}
A=\int d\r d\p f_0(\r,\p)\,,
\eeq
where $A$ is the number of particles.
This integral should keep the same value also out of equilibrium (global particle-number conservation).

\subsection {Constant-$\Delta$ approximation}

In a fully self-consistent approach, the pairing field $\Delta(\r,\p,t)$ is related to $\kappa(\r,\p,t)$, however here we introduce an approximation and replace the pairing field of the HFB theory with the phenomenological pairing gap of nuclei, hence in all our equations we put
\beq
\Delta(\r,\p,t)\approx \Delta_0(\r,\p)\approx\Delta=const.\,,
\eeq
with $\Delta\approx 1\MeV$. 

In the constant-$\Delta$ approximation the equilibrium distributions become
\bea
\label{rozero}
\rho_0(\epsilon)&=&\frac{1}{2}{\Big (}1-\frac{\epsilon-\mu}{E(\epsilon)}{\Big )}\,,\\
\label{kazero}
\kappa_0(\epsilon)&=&-\frac{\Delta}{2 E(\epsilon)}\,
\eea
and the quasiparticle energy
\bea
E(\epsilon)&=&\sqrt{\Delta^2+(\epsilon-\mu)^2}\,,
\eea
with $$\epsilon=h_0(\r,\p)=\frac{\p^2}{2m}+V_0(\r)$$ the particle energy in the equilibrium mean field. 

In the following we shall use the relation:
\beq
\label{deriv}
\kappa_0(\epsilon)=\frac{E^2(\epsilon)}{\Delta}\frac{d\rho_0(\epsilon)}{d\epsilon}\,.
\eeq

\section{Dynamic equations}

Always in  the approximation where $\Delta$ is constant and real, the time-dependent equations (\ref{even}--\ref{kappa})  become 
\bea
\label{even3}
i\hbar\partial_t\rho_{ev}&=&i\hbar\{h,\rho_{od}\}-2i\Delta{\rm Im}(\kappa)\,\\
\label{odd3}
i\hbar\partial_t\rho_{od}&=&i\hbar\{h,\rho_{ev}\}\,\\
\label{kappa3}
i\hbar\partial_t\kappa&=&2(h-\mu)\kappa-\Delta(2\rho_{ev}-1)\,.
\eea

This is the simplified set of equations that we want to study here. The sum of the first two equations gives an equation that is similar to the Vlasov equation of normal systems, only with the extra term 
$-2i\Delta{\rm Im}(\kappa)$. This extra term couples the equation of motion of $\rho$ with that of $\kappa$, thus, instead of a single differential equation (Vlasov equation), now we have a system of  two coupled differential equations (for $\rho$ and $\kappa_i$).

Our aim here is that of determining the effects of pairing on the linear response of nuclei, thus we
assume that our system is initially at equilibrium, with densities given by Eqs. (\ref{rozero},\ref{kazero}), and that at time $t=0$ a weak external field of the kind
\beq
\label{ext}
\delta V^{ext}(\r,t)=\beta\delta(t) Q(\r)
\eeq
is applied to it. This simple time-dependence is sufficient to determine the linear response of the system.
In a self-consistent approach, we should take into account also the changes of the mean field surrounding each particle induced by the external force and consider a perturbing hamiltonian of the kind
\beq
\delta h=\delta V^{ext}+\delta V^{int}\,,
\eeq
however we start with the zero-order approximation
\beq
\delta h=\delta V^{ext}
\eeq
and will consider collective effects in a second stage.

Since we want to solve Eqs. (\ref{even3}--\ref{kappa3})  in  linear approximation,  we consider  small fluctuations of the time-dependent quantities about their equilibrium values and neglect terms that are of second order in the fluctuations.
Hence, in Eqs. (\ref{even3}--\ref{kappa3}) we put:
\bea
h&=&h_0+\delta h\,,\\
\rho_{ev}&=&\rho_0+\delta\rho_{ev}\,,\\
\rho_{od}&=&\delta\rho_{od}\,,\\
\kappa&=&\kappa_0+\delta\kappa=\kappa_0+\delta\kappa_r+i\delta\kappa_i\,.
\eea
Then, the linearized form of Eqs. (\ref{even3}--\ref{kappa3}) is
\bea
\label{fast1}
i\hbar\partial_t\delta \rho_{ev}&=&i\hbar\{h_0,\delta\rho_{od}\}-2i\Delta\delta\kappa_i\,\\
\label{fast2}
i\hbar\partial_t\delta\rho_{od}&=&i\hbar\{h_0,\delta\rho_{ev}\}+i\hbar\{\delta h,\rho_0\}\,\\
\label{fast3}
-\hbar\partial_t\delta\kappa_i&=&2(\epsilon-\mu)\delta\kappa_r+2\kappa_0\delta h-2\Delta\delta\rho_{ev}\,,\\
\label{fast4}
\hbar\partial_t\delta\kappa_r&=&2(\epsilon-\mu)\delta\kappa_i\,.
\eea

Taking the sum of the first two equations gives
\bea
\label{last}
&&i\hbar\partial_t\delta\rho(\r,\p,t)=\\
&&i\hbar\{h_0,\delta\rho(\r,\p,t)\}+i\hbar\{\delta h(\r,\p,t),\rho_0\}-2i\Delta\delta\kappa_i(\r,\p,t)\,,\nonumber
\eea
which  can be regarded as an extension of the linearized Vlasov equation studied in \cite{bdd}. 

In order to make the comparison with \cite{bdd} easier, from now on we change the normalization of the phase-space densities and define
\bea
f(\r,\p,t)&=&\frac{4}{(2\pi\hbar)^3}\rho(\r,\p,t)\\
\chi(\r,\p,t)&=&\frac{4}{(2\pi\hbar)^3}\kappa(\r,\p,t)\,,
\eea
moreover, we put
\beq
F(\epsilon)=\frac{4}{(2\pi\hbar)^3}\rho_0(\epsilon)\,.
\eeq
In terms of the new functions Eqs. (\ref{last}) and (\ref{fast3}) read
\bea
\label{prima}
i\hbar\partial_t\delta f(\r,\p,t)&=&i\hbar\{h_0,\delta f(\r,\p,t)\}+i\hbar\{\delta h(\r,\p,t),f_0\}\nonumber\\
&-&2i\Delta\delta\chi_i(\r,\p,t)\,,\\
\label{seconda}
-\hbar\partial_t\delta\chi_i(\r,\p,t)&=&2(\epsilon-\mu)\delta\chi_r(\r,\p,t)
+2\chi_0\delta h(\r,\p,t)\nonumber\\
&-&2\Delta\delta f_{ev}(\r,\p,t)\,.
\eea
The function $f_{ev}$ is given by the obvious extension of Eq. (\ref{roev}).
In order to get a closed system of equations, we still need an extra equation for $\delta\chi_r(\r,\p,t)$. This can be obtained from the linearized form of the supplementary condition (\ref{scpau2}) that reads
\beq
\delta\rho_{ev}(2\rho_0-1)=-2\kappa_0\delta\kappa_r\,,
\eeq
or
\beq
\delta\kappa_r(\r,\p,t)=\frac{1-2\rho_0(\epsilon)}{2\kappa_0}\delta\rho_{ev}(\r,\p,t)=-\frac{\epsilon-\mu}{\Delta}\delta\rho_{ev}(\r,\p,t)\,.
\eeq
The last expression has been obtained with the help of Eq. (\ref{kappa2}).
In terms of the new functions $f$ and $\chi$, the last equation reads
\beq
\label{terza}
\delta\chi_r(\r,\p,t)=-\frac{\epsilon-\mu}{\Delta}\delta f_{ev}(\r,\p,t)\,.
\eeq

Equations (\ref{prima}, \ref{seconda}) and (\ref{terza}) are the set of coupled equations for the phase-space densities that we have to solve. 

Replacing Eq. (\ref{terza}) into Eq. (\ref{seconda}),  and using Eq. (\ref{kazero}), gives the following system of  coupled differential equations:
\bea
\partial_t\delta f(\r,\p,t)&=&\{h_0,\delta f\}+\{\delta h, f_0\}-2\frac{\Delta}{\hbar}\delta\chi_i(\r,\p,t)\,,\\
\partial_t\delta\chi_i(\r,\p,t)&=&\frac{E^2(\epsilon)}{\hbar\Delta}[\delta f(\r,\p,t)+\delta f(\r,-\p,t)]-2\frac{\chi_0}{\hbar}\delta h(\r,\p,t)\,.
\eea
Taking the Fourier transform in time, gives
\bea
-i\omega\delta f(\r,\p,\omega)&=&\{h_0,\delta f\}+\{\delta h, f_0\}-2\frac{\Delta}{\hbar}\delta\chi_i(\r,\p,\omega)\,,\\
-i\omega\delta\chi_i(\r,\p,\omega))&=&\frac{E^2(\epsilon)}{\hbar\Delta}[\delta f(\r,\p,\omega)+\delta f(\r,-\p,\omega)]\nonumber\\
&-&2\frac{\chi_0}{\hbar}\delta h(\r,\p,\omega)\,,
\eea
or (for $\omega\neq 0$)
\bea
\label{val}
-i\omega \delta f({\bf r},{\bf p},\omega) +\{\delta f, h_{0}\}
&=& -i\omega d^2\half[\delta f(\r,\p,\omega)+\delta f(\r,-\p,\omega)]\nonumber\\
&+& F'(\epsilon)[\{\delta h,h_0\}+i\omega d^2\delta h]\,,
\eea
with 
\beq
d^{2}={\Big (}\frac{\Omega(\epsilon)}{\omega}{\Big )}^2
\eeq
and
\beq
\Omega(\epsilon)=2\frac{E(\epsilon)}{\hbar}\,.
\eeq
This frequency plays a crucial role in our approach, its minimum value  is $2\Delta/\hbar$.

 In Eq. (\ref{val}) we have used the relation $\{f_0,\delta h\}=F'(\epsilon)\{h_o,\delta h\}$ as well as Eq. (\ref{deriv}).

By comparing Eq. (\ref{val}) with the analogous equation for normal systems
\beq
\label{lve}
-i\omega\delta f(\r,\p,\omega)+\{\delta f,h_0\}=F'(h_0)\{\delta h,h_0\}\,,
\eeq
we can see that the only effect of pairing in the constant-$\Delta$ approximation is that of adding the terms proportional to $d^2$.

The normal Vlasov equation (\ref{lve}) can be solved in a very compact way by using the method of action-angle variables  \cite{ps}, \cite{bdd}. In that  approach one expands
\beq
\delta h=\sum_{\n}\delta h_{\n}(\bf I)e^{i\n\cdot {\bf \Phi}}\,,
\eeq
where $\bf I$ and $\bf \Phi$ are the action and angle variables, respectively.
Moreover
\beq
\label{delf}
\delta f(\r,\p,\omega)=\sum_\n\delta f_\n({\I},\omega)e^{i\n\cdot {\bf \Phi}}
\eeq
and
\beq
\{\delta f,h_0\}=\sum_\n i(\n\cdot\vec{\omega})\delta f_\n({\I},\omega)e^{i\n\cdot {\bf \Phi}}\,,
\eeq
where   the vector $\vec{\omega}$ has components
\beq
\omega_\alpha=\frac{\partial h_0}{\partial I_\alpha}\,.
\eeq
Then Eq.(\ref{lve}) gives
\beq
\delta f_\n({\I},\omega)=\frac{(\n\cdot\vec{\omega})}{(\n\cdot\vec{\omega})-(\omega+i\varepsilon)}F'(\epsilon)\delta h_{\n}(\bf I)\,.
\eeq
 The (zero-order) eigenfrequencies of the (normal) physical system are
 \beq
 \label{uncoef}
 \omega_{\bf n}={\bf n}\cdot\vec{\omega}\,.
 \eeq
 Here want to use the same method to solve the more complicated equation (\ref{val}). Since that equation contains also the function $\delta f(\r,-\p,\omega)$, we need also the analogous equation for this other quantity:
\bea
\label{val2}
-i\omega \delta f({\bf r},-{\bf p},\omega) -\{\delta f, h_{0}\}_{\r,\p}
&=& -i\omega d^2\half[\delta f(\r,\p,\omega)+\delta f(\r,-\p,\omega)]\nonumber\\
&+& F'(\epsilon)[-\{\delta h,h_0\}_{\r,\p}+i\omega d^2\delta h]\,.
\eea

By expanding $\delta f(\r,\p,\omega)$ and $\delta f(\r,-\p,\omega)$ as
\beq
\label{deltafpm}
\delta f(\r,\pm\p,\omega)=\sum_\n\delta  f^\pm_\n({\I},\omega)e^{i\n\cdot {\bf \Phi}}\,,
\eeq
Eqs. (\ref{val}) and (\ref{val2}) give
\bea
&&[-\omega(1-\frac{d^2}{2})+\omega_\n]\delta f^+_\n+\omega\frac{d^2}{2}\delta  f^-_\n=F'(\epsilon)[\omega_\n+\omega d^2]\delta h_\n\,,\quad\\
&&[\omega\frac{d^2}{2}]\delta f^+_\n+[-\omega(1-\frac{d^2}{2})-\omega_\n]\delta  f^-_\n=F'(\epsilon)[-\omega_\n+\omega d^2]\delta h_\n\,,\quad
\eea
which is a system of two coupled algebraic equations for the coefficients $\delta f^+_\n$ and $\delta f^-_\n$. Its solution is
\bea
\label{sol1}
&&\delta f^+_{\bf n}=\frac{\bar\omega^2_{\bf n}+\omega\omega_{\bf n}}{\bar\omega^2_{\bf n}-\omega^2}F'(\epsilon)\delta h_{\bf n}\,,\\
\label{sol2}
&&\delta  f^-_{\bf n}=\frac{\bar\omega^2_{\bf n}-\omega\omega_{\bf n}}{\bar\omega^2_{\bf n}-\omega^2}F'(\epsilon)\delta h_{\bf n}\,,
\eea
where
\beq
\label{coef}
\bar\omega^2_{\bf n}=\omega^2_{\bf n}+\Omega^2(\epsilon)
\eeq
are the (squared) eigenfrequencies of the correlated system. These eigenfrequencies are in agreement with the enegy spectrum of a superfluid infinite homogeneous Fermi gas (see e. g. Sect. 39 of \cite{ll9}) and they lead to a low-energy gap of $2\Delta$ in the excitation spectrum of the correlated systems. However, as anticipated, we expect problems with particle-number conservation. These problems are better discussed in one dimension, where formulae are simpler. 

\section{One-dimensional systems and particle-number conservation}
\label{oned}

In one dimension, Eq.(\ref{val}) reads
\bea
&&-i\omega \delta f(x,p,\omega)+ \dot x\partial_x\delta f-\frac{dV_0(x)}{dx}\partial_{p} \delta f=\\
&&-i\omega d^2\frac{1}{2}[\delta f(x,p,\omega)+
\delta f(x,-p,\omega)]+F'(\epsilon)(\frac{p}{m}\partial_x\delta h+i\omega d^2\delta h)\,.\nonumber
\eea

In zero-order approximation $\delta h(x,p,\omega)=\beta Q(x)$, moreover, in one dimension
\beq
F'(\epsilon)=\frac{4}{2\pi\hbar}\frac{d\rho_0}{d\epsilon}\,.
\eeq
The vectors  $\vec{\omega}$ and ${\bf \Phi}$ have only one component: 
\beq
\omega_0(\epsilon)=\frac{2\pi}{T(\epsilon)}
\eeq
and
\beq
\Phi(x)=\omega_0\tau(x)\,,
\eeq
with
\bea
&&\tau(x)=\int_{x_1}^x dx'\frac{1}{v(\epsilon,x')}\,,\\
&&v(\epsilon,x)=\sqrt{\frac{2}{m}[\epsilon-V_0(x)]}\,.
\eea
The time $T(\epsilon)$ is the period of the bound motion of particles with enegy $\epsilon$ in the equilibrium potential well $V_0(x)$: $T=2\tau(x_2)$. The points $x_{1,2}$ are the classical turning points for the same particles.  Instead of the action variable $I(\epsilon)=\frac{1}{2\pi}\oint dx p(\epsilon,x)$, it is more convenient to use the particle energy $\epsilon$ as  constant of  motion. 
As pointed out in \cite{bdd}, the range of values of $\tau$ can be extended to the whole interval $(0,T)$, by defining
\beq
\tau(x)=\frac{T}{2}+\int_x^{x_2} dx'\frac{1}{v(\epsilon,x')}\,
\eeq
when $\tau>\frac{T}{2}$.
 With this extension, the angle variable $\Phi(x)$ takes values between $0$ and $2\pi$, as it should.

In one dimension,  Eqs. (\ref{sol1}, \ref{sol2}) give
\beq
\label{sol1/1}
\delta f^\pm_n(\epsilon,\omega)=\frac{\bar\omega^2_n\pm\omega\, \omega_n}{\bar\omega^2_n-\omega^2}F'(\epsilon)\delta h_n\,,
\eeq
with
\bea
\label{fco}
&&\delta h_n=\beta Q_n\nonumber\\
&&=\beta\frac{1}{T}\oint dx\frac{Q(x)}{v(\epsilon,x)}e^{-i\omega_n\tau(x)}\nonumber\\
&&=\beta\frac{2}{T}\int_{x_1}^{x_2} dx\frac{Q(x)}{v(\epsilon,x)}\cos[\omega_n\tau(x)]\,,
\eea
The frequencies $\omega_n$ are the eigenfrequencies of the uncorrelated system:
\beq
\omega_n=n\omega_0\,,
\eeq
while $\bar\omega_n$ are the new eigenfrequencies modified by the pairing correlations:
\beq
\bar\omega_n=\pm\sqrt{\omega_n^2+\Omega^2(\epsilon)}\,.
\eeq

 Note that, since $\delta h_{-n}=\delta h_n$, then $\delta  f^-_n=\delta f^+_{-n}$.
 
 By using the solutions (\ref{sol1/1}), we can also obtain an expansion for the even and odd parts
of $\delta f$:
\bea
\label{fev}
&&\delta f_{ev}(x,\epsilon,\omega)=\sum_{ n=0}^{\infty}A_{ n}(\omega)\cos n\omega_0\tau(x)\,,\\
\label{fod}
&&\delta f_{od}(x,\epsilon,\omega)=\sum_{ n=1}^{\infty}B_{ n}(\omega)\sin n\omega_0\tau(x)\,,
\eea
with
\bea
\label{an}
&&A_{n}(\omega)=\frac{\bar\omega^2_{ n}}{\bar\omega^2_{n}-\omega^2}F'(\epsilon)\delta h'_{ n}\,,\\
&&B_{n}(\omega)=i\omega\frac{\omega_n}{\bar\omega^2_{ n}-\omega^2}F'(\epsilon)\delta h'_{ n}\,
\eea
and
\bea
\delta h'_n&=&2\delta h_n\,,\qquad\qquad\:\; n\neq0\,,\\
&=&\delta h_n\,,\qquad\qquad\quad  n=0\,.
\eea
Note that, while $B_{n=0}(\omega)=0$, we have $A_{n=0}(\omega)\neq 0$, and this fact leads to an unphysical fluctuation of the number of particles,  induced by the applied external field. These fluctuations are given by
\beq
\delta A(\omega)=\int d x\delta \varrho(x,\omega)\,,
\eeq
where $\delta\varrho(x,\omega)$ is the density fluctuation at point $x$:
\beq
\label{wrong}
\delta\varrho(x,\omega)=\int d p\;\delta f(x,p,\omega)=2\int\frac{d\epsilon}{v(\epsilon,x)}\delta f_{ev}(x,\epsilon,\omega)\,.
\eeq
Equation (\ref{fev}) gives
\beq
\delta A(\omega)=2\sum_{n=0}^\infty A_n(\omega)\int_0^{T/2}d\tau\cos n\omega_0\tau\,.
\eeq
Since the integrals $\int_0^{T/2}d\tau\cos n\omega_0\tau$ vanish when $n\neq 0$, the term with $n=0$
is the only one contributing to this sum, thus givig an unphysical  fluctuation of the number of particles.
This problem could be solved simply by excluding the mode $n=0$ from the sum in Eq. (\ref{fev}), however this would  not be sufficient to solve all problems with  particle-number conservation, since we can easily check that the solutions (\ref{fev}, \ref{fod}) do not satisfy the continuity equation
\beq
\label{cont}
i\omega\varrho(x,\omega)=\partial_x j(x,\omega)\,.
\eeq

The density fluctuation involves only the even part of $\delta f$, while the current density $j(x,\omega)$ involves only the odd part:
\beq
j(x,\omega)=\int dp\frac{p}{m}\delta f(x,p,\omega)=2\int d\epsilon\delta f_{od}(x,\epsilon,\omega)\,.
\eeq
The fact that the continuity equation is violated is a very serious shortcoming of the constant-$\Delta$ approximation. However, since we have seen that this approximation leads to very simple equations and to rather satisfactory expressions for the eigenfrequencies of the correlated systems, we still use it, but with the following prescription: {\em when calculating the longitudinal response function, the density fluctuations should be evaluated by using Eq. (\ref{cont}), instead of Eq, (\ref{wrong})}.
Then, the density fluctuations (\ref{wrong}) should be replaced by
\beq
\label{barro}
\delta\bar\varrho(x,\omega)=\frac{2}{i\omega}\int d\epsilon\partial_x\delta f_{od}(x,\epsilon,\omega)\,.
\eeq

In practice we are proposing to evaluate the longitudinal response function in terms of the transverse response function. It is well known that also the more familiar BCS approximation gives a more accurate description of the transverse response (see e.g. sect. 8-5 of \cite{sch}).
In the Appendix we show that the  longitudinal response function resulting from the present  prescription satisfies the same energy-weighted sum rule as the uncorrelated response function. This would not necessarily happen if, instead of changing only the even part of $\delta f$, we had modified also its odd part.

It is interesting to see how the solutions (\ref{sol1/1}) are changed by our prescription.
By using Eq. (\ref{fod}) for$f_{od}$, Eq. (\ref{barro}) gives
\beq
\delta\bar\varrho(x,\omega)=2\int \frac{d\epsilon}{v(\epsilon,x)}\delta \bar f_{ev}(x,\epsilon,\omega)\,,
\eeq
with
\beq
\delta \bar f_{ev}(x,\epsilon,\omega)=\sum _{n=0}^\infty\bar A_n(\omega)\cos n\omega_0\tau(x)\,,
\eeq
and
\beq
\bar A_n(\omega)=\frac{\omega_n}{i\omega}B_n(\omega)\,,
\eeq
(note that $\bar A_{n=0}(\omega)=0$). Then
\beq
\delta \bar f(x,\pm p,\omega)=\delta\bar f_{ev}(x,\epsilon,\omega)\pm\delta f_{od}(x,\epsilon,\omega)\,
\eeq
and Eq. (\ref{sol1/1}) is replaced by
\beq
\delta \bar f^\pm_n(\epsilon,\omega)=\frac{\omega^2_n\pm\omega\, \omega_n}{\bar\omega^2_n-\omega^2}F'(\epsilon)\delta h_n\,.
\eeq
By comparing this expression to Eq. (\ref{sol1/1}), we can see that the fluctuations of the phase-space density given by the constant-$\Delta$ approximation contain an extra  contribution that we identify as spurious:
\beq
\delta f_n^\pm(\epsilon,\omega)=\delta \bar f^\pm_n(\epsilon,\omega)+\delta f^{spur}_n(\epsilon,\omega)\,,
\eeq
with
\beq
\label{spur}
\delta f^{spur}_n(\epsilon,\omega)=\frac{\Omega^2(\epsilon)}{\bar\omega_n^2-\omega^2}F'(\epsilon)\delta h_n\,.
\eeq
The spurious character of $\delta f^{spur}_n$ is suggested also by sum-rule arguments (see Appendix). The term $f^{spur}_n(\epsilon,\omega)$ contributes to all modes of the density strength function: the contribution to the mode $n=0$ gives a fluctuation of the particle-number integral (global paticle-number violation), while the other modes give a spurious contribution to the density strength function, increasing the sum rule and violating the continuity equation (local particle-number violation). Note that the spurious contribution
(\ref{spur}) affects only the even part of the pase-space density, not the odd part.

\section{Spherical Systems}

The method of action-angle variables gives a very compact solution of the linearized Vlasov equation both in the uncorrelated and correlated cases, however it may be useful to make a connection between
the results given by this method and the more explicit treatment of spherical nuclei given in \cite{bdd}.
For uncorrelated system this has been done in  \cite{dmb}. Here we follow that approach in order to derive  useful expressions for correlated spherical systems. The components of the vecor ${\bf n}$ are $(n_1,n_2,n_3)$, the first point to notice is that, because of the degeneracy associated with any central-force field, the vector $\vec{\omega}$ has only two non vanishing components:
\beq
\vec{\omega}=(0,\omega_\varphi(\epsilon,\lambda), \omega_0(\epsilon,\lambda))\,.
\eeq
With $\lambda$ we denote the magnitude of the particle angular momentum. According to Eq. (\ref{uncoef}), the eigenfrequencies of the uncorrelated system are \cite{bdd}
\beq
\omega_{\bf n}=\omega_{n_3,n_2}(\epsilon,\lambda)=n_3\omega_0+n_2\omega_\varphi\,,
\eeq
while Eq. (\ref{coef}) gives the  correlated eigenfrequencies
\beq
\bar\omega_{\bf n}=\bar\omega_{n_3,n_2}=\pm\sqrt{\omega^2_{n_3,n_2}+\Omega^2(\epsilon)}\,.
\eeq

In three dimensions, the Fourier coefficients analogous to (\ref{fco}) are
\beq
\label{3fc}
Q_{\bf n}({\bf I})=\frac{1}{(2\pi)^3}\int d{\bf \Phi} e^{-i{\bf n}\cdot{\bf \Phi}} Q(\r)\,.
\eeq
The external field $Q(\r)$ can be expanded in partial waves as
\beq
Q(\r)=\sum_{LM} Q_{L}(r) Y_{LM}(\hat\r)\,,
\eeq
giving
\beq
\label{exp}
Q_{\bf n}({\bf I})=\sum_{LM}Q_{\bf n}^{(LM)}\,,
\eeq
with \cite{dmb}
\beq
Q_{\bf n}^{(LM)}=\sum_{N=-L}^LY_{LN}{\Big (}\frac{\pi}{2},0{\Big)}d^L_{MN}(\beta')\delta_{M,n_1}\delta_{N,n_2}Q^L_{n_3N}\,.
\eeq

By using this last equation (and changing $n_3\to n$), the expansion (\ref{deltafpm}) becomes
\bea
\label{sol1/3}
&&\delta f(\r,\pm\p,\omega)=\sum_{L=0}^\infty\quad\sum_{M=-L}^L\quad\sum_{N=-L}^L\quad\sum_{n=-\infty}^\infty\\
&&\delta f^{L\pm}_{nN}(\epsilon,\lambda,\omega)e^{i\phi_{nN}(r)}{\Big(}{\cal D}^L_{MN}(\alpha,\beta',\gamma){\Big)}^*Y_{LN}{\Big (}\frac{\pi}{2},\frac{\pi}{2}{\Big)}\nonumber\,,
\eea
with
\beq
\label{coefplus}
 \delta f^{L\pm}_{nN}(\epsilon,\lambda,\omega)=
\frac{\bar\omega^2_{nN}\pm\omega\omega_{nN}}{\bar\omega^2_{nN}-\omega^2}\beta F'(\epsilon)Q^L_{nN}\,
\eeq
and $Q^L_{nN}$  the semiclassical limit of the  radial matrix elements:
\bea
Q^L_{nN}&=&\frac{1}{T}\oint \frac{d r}{v_r(r)}e^{-i\phi_{nN}(r)}Q_{L}(r)\,,\nonumber\\
&=&\frac{2}{T}\int_{r_1}^{r_2}\frac{d r}{v_r(r)}\cos[\phi_{nN}(r)]Q_{L}(r)\,.
\eea
Here $T$ is the period of radial motion, $v_r(r)$ the radial velocity
\beq
v_r(r)=\sqrt{\frac{2}{m}{\Big(}\epsilon-V_0(r)-\frac{\lambda^2}{2mr^2}{\Big)}}
\eeq
and the phases $\phi_{nN}(r)$ are given by
\beq
\phi_{nN}(r)=\omega_{nN}\tau(r)-N\gamma(r)\,,
\eeq
where
\beq
\tau(r)=\int_{r_1}^r\frac{dr'}{v_r(r')}
\eeq
and
\beq
\gamma(r)=\int_{r_1}^r\frac{dr'}{v_r(r')}\frac{\lambda}{mr'^2}\,.
\eeq

The frequencies $\omega_0$ and $\omega_\varphi$ are given by
\bea
\omega_0&=&\frac{\pi}{\tau(r_2)}\,,\\
\omega_\varphi&=&\frac{\gamma(r_2)}{\tau(r_2)}\,.
\eea
The Wigner rotation matrix elements in Eq. (\ref{sol1/3}) are given by \cite{bsa}
\beq
{\cal D}^L_{MN}(\alpha,\beta',\gamma)=e^{-iM\alpha}d^L_{MN}(\beta')e^{-iN\gamma}\,,
\eeq
where $(\alpha,\beta',\gamma)$ are the Euler angles introduced in \cite{bdd}.

On the basis of the discussion in Sect. (\ref{oned}), we expect that the solution (\ref{sol1/3}) will contain some spurious strength introduced by the constant-$\Delta$ approximation. In order to eliminate the spurious contributions, we should replace the coefficients (\ref{coefplus}) with
\beq
\delta \bar f^{L\pm}_{nN}(\epsilon,\lambda,\omega)=\frac{\omega^2_{nN}\pm
\omega\omega_{nN}}{\bar\omega^2_{nN}-\omega^2}\beta F'(\epsilon)Q^L_{nN}\,.
\eeq

These modified coefficients allow us to obtain the modified zero-order propagator
\beq
\label{cprop}
\bar D^0_L(r,r',\omega)=\int d\epsilon F'(\epsilon) \int d\lambda\,\lambda\sum_{n=-\infty}^\infty\quad\sum_{N=-L}^L\frac{\bar d^L_{nN}(r,r')}{\omega-\bar\omega_{nN}+i\varepsilon}\,,
\eeq
with
\bea
\label{dln}
&&\bar d^L_{nN}(r,r')=\\
&&\frac{(4\pi)^2}{2L+1}|Y_{LN}(\frac{\pi}{2},\frac{\pi}{2})|^2{\Big (}\frac{-2\omega_{nN}}{T}{\Big)}{\Big(}\frac{\omega_{nN}}{\bar\omega_{nN}}{\Big)}\frac{\cos\phi_{nN}(r)}{r^2v_r(r)}\frac{\cos\phi_{nN}(r')}{r'^2v_r(r')}\nonumber
\eea
and the corresponding  response and strength functions:
\bea
\bar{\cal R}^0_L(\omega)&=&\int dr dr' r^2Q_L(r)\bar D^0_L(r,r',\omega)r'^2Q_L(r')\,,\\
\bar S^0_L(\omega)&=&-\frac{1}{\pi}{\rm Im}\bar{\cal R}^0_L(\omega)\,.
\eea
For multipole response: $Q_L(r)=r^L$.

For normal systems, the zero-order propagator $D^0_L(r,r',\omega)$ is given by Eqs. (\ref{cprop}) and(\ref{dln}) where $\bar\omega_{nN}$ is replaced by $\omega_{nN}$ and $F'(\epsilon)$ is proportional to a $\delta$-function\cite{bdd}.

\section{Collective response}

Up to now, we have been concerned only with the zero-order approximation, which corresponds to the single-particle approximation of the quantum approach. In this approximation, the perturbing part of the hamiltonian is given only by the external field, while a more consistent approach would require taking into account also the mean-field fluctuation induced by the external force, so that
\beq
\delta h=\delta V^{ext}(\r,\omega)+\delta V^{int}(\r,\omega)\,.
\eeq
In the Hartree approximation,
\beq
\delta V^{int}(\r,\omega)=\int d\r' v(\r-\r')\delta\varrho(\r',\omega)\,.
\eeq
where $v(\r-\r')$is the (long-range) interaction between constituents.

For consistency, we take
\beq
\delta V^{int}(\r,\omega)=\int d\r' v(\r-\r')\delta\bar\varrho(\r',\omega)\,,
\eeq
then the collective propagator for correlated systems satisfies the same kind of integral equation as for normal systems \cite{bdd}:
\bea
\label{rpa}
&&\bar D_L(r,r',\omega)=\\
&&\bar D^0_L(r,r',\omega)+\int dx x^2\int dy y^2 \bar D^0_L(r,x,\omega)v_L(x,y)\bar D_L(y,r',\omega)\,.\nonumber
\eea
Here $v_L(x,y)$ is the partial-wave component of the interaction between particles. We assume that this interaction can be approximated by a separable form of the kind
\beq
v_L(x,y)=\kappa_Lx^Ly^L\,,
\eeq
where $\kappa_L$ is a parameter that determines the strength of the interaction. Then, the integral equation (\ref{rpa}) gives an algebraic equation for the collective correlated response function $$\bar{\cal R}_L(\omega)=\int dr r^2 r^L\int dr'r'^2 r'^L \bar D_L(r,r',\omega)$$ leading to the expression
\beq
\label{coll}
\bar{\cal R}_L(\omega)=\frac{\bar{\cal R}^0_L(\omega)}{1-\kappa_L\bar{\cal R}^0_L(\omega)}\,.
\eeq
%\newpage
\section{Results}

Here we compare the multipole strength functions given by our simplified model of pairing correlations with that of the corresponding uncorrelated system. This comparison is made for the quadrupole and octupole strength functions, since these channels are the ones that are most affected. 

The static nuclear mean field is approximated with a spherical cavity of radius $R=1.2\,A^\frac{1}{3}$ fm and the $A$ nucleons are treated on the same footing, i. e., we do not distinguish between neutrons and protons. Moreover, we chose $A=208$ for ease of comparison with previous calculations of uncorrelated response functions \cite{adm,addm}.  Shell effects are not included in our semiclassical picture and the results shown  below should be considered  as an indication of the qualitative effects to be expected in heavy nuclei.
For the uncorrelated calculations, the Fermi energy is determined by the parametrization chosen for the radius as $\epsilon_F\approx 33.33$ MeV, while for the correlated case, the parameter $\mu$ is determined by the condition (\ref{norm}); with the value of $\Delta=1$ MeV used here, the value of $\mu$ is practically coincident with that of $\epsilon_F$, so we have used $\mu=\epsilon_F=33.33$ MeV in the calculations   below. Moreover, the small parameter $\varepsilon$ appearing in Eq. (\ref{cprop}) has been given the  value $\varepsilon=0.1$ MeV. This value is chosen to simplify the evaluation of the response function by smoothing out discontinuities in its dependence on $\omega$.

In the evaluation of the collective response, the value of parameters $\kappa_L$ is the same as in  \cite{adm,addm}, that is: $\kappa_2=-1\times10^{-3}\;{\rm MeV/fm^4}$ and $\kappa_3=-2\times10^{-5}\;{\rm MeV/fm^6}$.

\subsection{Quadrupole response}

\begin{figure}[h]
\vspace{.2in}
\centerline {
\includegraphics[width=1.8in]{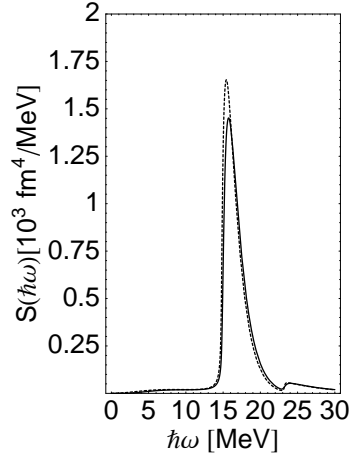}
}
\vspace{.2in}
\caption{Quadrupole stength function in zero-order approximation. The dashed curve gives the response of a normal system of $A=208$ nucleons contained in a spherical cavity, while the solid curve includes the effects of pairing correlations in constant-$\Delta$ approximation.}
\end{figure}
\begin{figure}[h]
\vspace{.2in}
\centerline {
\includegraphics[width=1.8in]{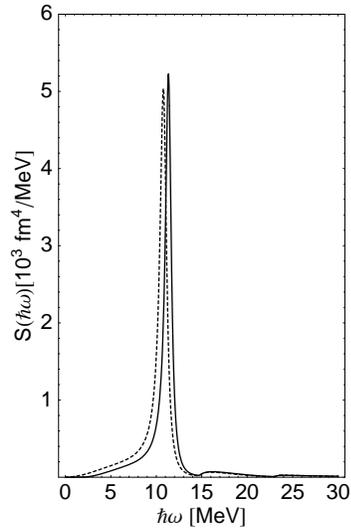}
}
\vspace{.2in}
\caption{Collective quadrupole strength function showing the giant quadrupole resonance. The solid curve involves also pair correlations,  the dashed curve has no pairing.}
\end{figure}
%\newpage
 \begin{figure}[h]
\vspace{.2in}
\centerline {
\includegraphics[width=2.8in]{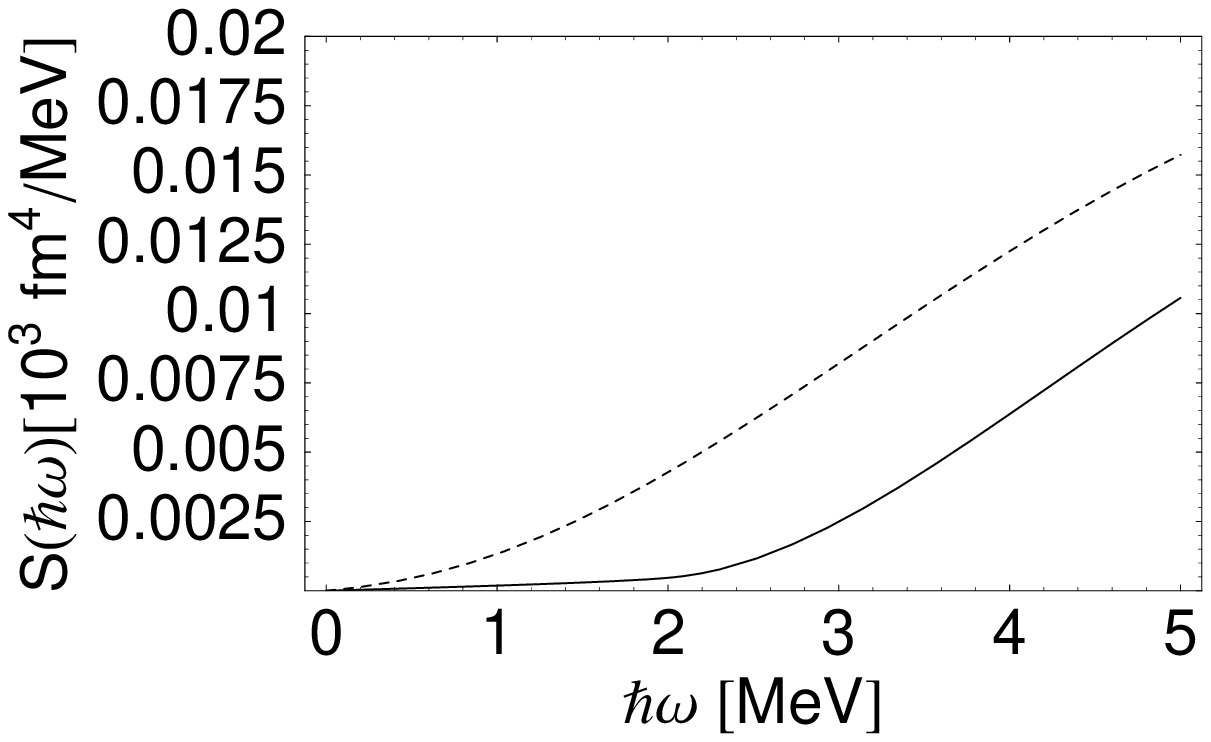}
}
\vspace{.2in}
\caption{Same as Fig.1, at low excitation energy.}
\end{figure}

Figure 1 show the longitudinal quadrupole strength function evaluated in the zero-order approximation (corresponding to the quantum single-particle approximation). The dashed curve shows the uncorrelated response evaluated according to the theory of \cite{bdd}, while the full curve shows the result of the present correlated calculation.  As we can see the effect of pairing correlations on this zero-order strength function is rather small, however, since pairing affects also the real part of the zero-order response function,in Fig. 2 we plot also the collective strength function given by Eq. (\ref{coll}). Again, the effect of pairing correlations is very small, in agreement with the results of \cite{dk}.

The main difference between the uncorrelated and correlated responses occurs at small excitation energy, Fig. 3 shows a detail of Fig. 1 at low excitation energy. The correlated strength function displays a gap of about  2 $\Delta$ , the very small strength extending below 2 MeV is entirely due to the finite value of the small parameter $\varepsilon$ used in the numerical evaluation of the propagator  (\ref{cprop}).
%\newpage
\subsection{Octupole response}

Figures 4 and 5 show the zero-order  and collective octupole strength functions, both correlated and uncorrelated. As we can see,  in this case too the effect is rather  small.
 \begin{figure}[h]
\vspace{.2in}
\centerline {
\includegraphics[width=1.8in]{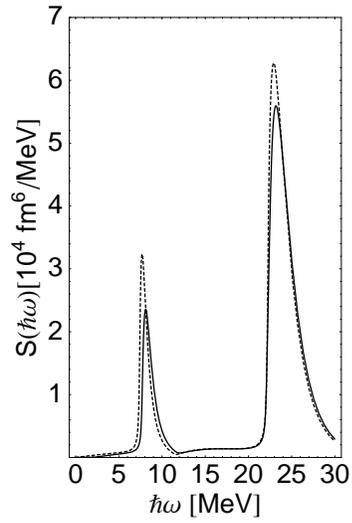}
}
\vspace{.2in}
\caption{Zero-order octupole strength function. The solid curve involves  pair correlations,  the dashed curve has no pairing.}
\end{figure}
%\newpage
 \begin{figure}[h]
\vspace{.2in}
\centerline {
\includegraphics[width=1.8in]{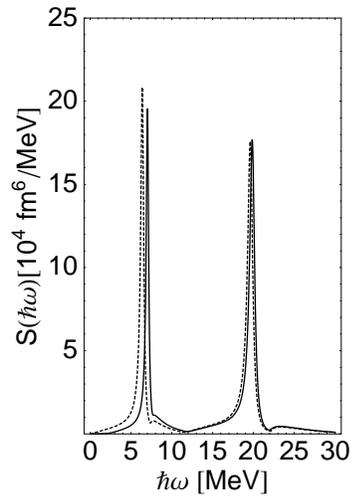}
}
\vspace{.2in}
\caption{Collective octupole strength function. The solid curve involves  pair correlations,  the dashed curve has no pairing.}
\end{figure}
\newpage
\section{Conclusions}

The solutions of the semiclassical time-dependent Hartree--Fock--Bogoliubov equations have been studied in a simplified model in which the pairing field $\Delta(\r,\p,t)$ is treated as a constant phenomenological parameter. Such an approximation is known to violate some important constraints, like global (particle-number integral) and local (continuity equation) particle-number conservation. In a linearized approach , we have shown that the global particle-number violation is related only to one particular mode of the density fluctuations, while the violation of the continuity equation gives a spurious contribution to all modes of the density response.  Both global and local   particle-number conservation can be restored by introducing a new density fluctuation that is related to the current density by the continuity equation. This prescription changes the strength associated with the various eigenmodes of the density fluctuations, but not the eigenfrequencies of the system. We have shown in a one-dimensional model that the energy-weighted sum rule calculated according to this prescription has exactly the same value as for normal, uncorrelated systems, thus we conclude that our prescription eliminates all  the spurious strength introduced by the constant-$\Delta$ approximation.

In a simplified model of nuclei, the effects of pairing correlations on the isoscalar strength functions has been studied in detail for the quadrupole and octupole channels, in the region of giant resonances. In both cases the effects of pairing  are rather small. More sizable effects are found at lower excitation energy, in the region of surface modes which have not been included in the present model, but will certainly be more affected by pairing correlations.

\section*{Appendix}

In this Appendix we show that the correlated zero-order response function given by the modified density fluctuation (\ref{barro})  satisfies the same energy-weighted sum rule (EWSR) as the uncorrelated response function. We assume that particles move in  one-dimensional square-well potential, so that formulae become simpler because the particle velocity does not depend on position: 
$v(\epsilon,x)=v(\epsilon)=\sqrt{\frac{2\epsilon}{m}}\,.$

\subsection*{Uncorrelated sum rule}

The uncorrelated propagator \cite{bdd}
\bea
\label{unco}
&&D^0(x,x',\omega)=\\
&&2\int d\epsilon F'(\epsilon)\sum_{n}\frac{-2n\omega_0}{T}\frac{\cos[n\omega_0\tau(x)]}{v(\epsilon,x)}\frac{1}{\omega-n\omega_0(\epsilon)+i\varepsilon}\frac{\cos[ n\omega_0\tau(x')]}{v(\epsilon,x')}\,\nonumber
\eea
gives the uncorrelated  strength function $ S^0(\omega)=-\frac{1}{\pi}{\rm Im}\int dx dx'Q(x)D^0(x,x',\omega)Q(x')$
and the first moment
\bea
\label{mom}
M_1&=&\int_0^\infty d\omega\omega S^0(\omega)\nonumber\\
&=&2\int_0^\infty d\epsilon F'(\epsilon)\sum_{n>0}{\Big (}\frac{-2n\omega_0(\epsilon)}{T(\epsilon)}{\Big )}
{\Big (}\frac{T(\epsilon)}{2}{\Big )}^2Q_n^2 n\omega_0(\epsilon).
\eea

\subsection*{Correlated sum rule}

The modified density fluctuation (\ref{barro}) allows us to evaluate the correlated propagator $\bar D^0$ through the relation
\beq
\delta\bar\varrho(x,\omega)=\beta\int dx'\bar D^0(x,x',\omega)Q(x')\,,
\eeq
giving
\bea
\label{cprop/1}
&&\bar D^0(x,x',\omega)=\\
&&2\int d\epsilon F'(\epsilon)\sum_n\frac{-2n\omega_0}{T}{\Big (}\frac{n\omega_0}{\bar\omega_n}{\Big )}\frac{\cos n\omega_0\tau(x)}{v(\epsilon,x)}\frac{1}{\omega-\bar\omega_n+i\varepsilon}\frac{\cos n\omega_0\tau(x')}{v(\epsilon,x')}\nonumber
\eea
and the correlated first moment
\beq
\label{cmom}
\bar M_1=2\int_0^\infty d\epsilon F'(\epsilon)\sum_{n>0}{\Big (}\frac{-2n\omega_0(\epsilon)}{T(\epsilon)}{\Big )}{\Big (}\frac{T(\epsilon)}{2}{\Big )}^2Q_n^2{\Big (}\frac{n\omega_0(\epsilon)}{\bar\omega_n(\epsilon)}{\Big )}\bar\omega_n(\epsilon)\,.
\eeq

The only difference between this expression and Eq. (\ref{mom}) is in the form of $F'(\epsilon)$, which is proportional to a  $\delta$-function in (\ref{mom}), while it is smoother in the correlated case, however, if the parameter $\mu$ is determined by the one-dimensional version of Eq. (\ref{norm}), then it can be easly found that, for a square-well mean field,
\beq
\label{sr}
\bar M_1=M_1\,.
\eeq
The detailed argument goes as follows: both for correlated and uncorrelated systems, the number of particles is given by
\beq
\label{pnumb}
A=\int dx dp F(\epsilon)=\int_0^\infty d\epsilon T(\epsilon) F(\epsilon)\,,
\eeq
with $F(\epsilon)=\frac{4}{2\pi\hbar}\theta (\epsilon_F-\epsilon)$ for uncorrelated fermions and
$F(\epsilon)=\frac{4}{2\pi\hbar}\rho_0(\epsilon)$ in the correlated case, while the moments (\ref{mom}, \ref{cmom}) are given by
\beq
M=\int_0^\infty d\epsilon F'(\epsilon) G(\epsilon)\,,
\eeq
with 
\beq
G(\epsilon)=-\frac{(2\pi)^2}{T(\epsilon)}\sum_{n>0}Q_n^2
\eeq
in both cases,  $F'(\epsilon)$ obviously differs in the two cases.

Integrating by parts the last expression in (\ref{pnumb}), gives
\beq
A=J(\epsilon)F(\epsilon){\Big |}_0^\infty-\int_0^\infty F'(\epsilon) J(\epsilon)\,,
\eeq 
with
\beq
J(\epsilon)=\int d\epsilon T(\epsilon)\,.
\eeq
For a square-well potential of size $L$:
\bea
T(\epsilon)&=&\frac{\sqrt{2m}L}{\sqrt{\epsilon}}\,,\\
J(\epsilon)&=&\sqrt{2m}L\,2\sqrt{\epsilon}\,.
\eea
Since
\beq
\lim_{\epsilon\to 0}J(\epsilon)F(\epsilon)=\lim_{\epsilon\to\infty}J(\epsilon)F(\epsilon)=0\,,
\eeq
both for the correlated and uncorrelated distributions, we have
\beq
A=-\int_0^\infty d\epsilon F'(\epsilon)J(\epsilon)\,.
\eeq
The explicit expressions of $J(\epsilon)$ and $T(\epsilon)$ give
\bea
A&=&-2L\sqrt{2m}\int_0^\infty d\epsilon F'(\epsilon)\sqrt{\epsilon}\,,\\
M&=&-\frac{(2\pi)^2}{\sqrt{2m}L}\sum_{n>0}Q_n^2\int_0^\infty d\epsilon F'(\epsilon)\sqrt{\epsilon}
\eea
for both distributions. From these relations follows  that, for a square-well mean field, the relation (\ref{sr}) is exact. If we had used the density fluctuation (\ref{wrong}), instead of (\ref{barro}), to evaluate the correlated propagator, we would have obtained a different value of the first moment because
the additional term (\ref{spur}) in the phase-space density gives an extra contribution to the density response function and hence to the EWSR. Because of the fundamental character of the EWSR, as well as of the continuity equation, we conclude that this term is a spurious contribution generated by the constant-$\Delta$ approximation.
\end{document}